\documentclass[12pt,preprint]{aastex}
\begin{document}

\title{ASCA Observations of the Absorption-Line Features from the
Super-Luminal Jet Source GRS 1915+105}
\author{Taro Kotani\altaffilmark{1}, Ken Ebisawa\altaffilmark{2},
Tadayasu Dotani\altaffilmark{2}, 
Hajime Inoue\altaffilmark{2}, Fumiaki Nagase\altaffilmark{2}, 
Yasuo Tanaka\altaffilmark{3},
Yoshihiro Ueda\altaffilmark{4}}
\authoremail{kotani@hp.phys.titech.ac.jp}

\altaffiltext{1}{Tokyo Tech, 2-12-1 O-okayama, Tokyo 152-8551, Japan}
\altaffiltext{2}{ISAS/JAXA, 3-1-1 Yoshinodai, Sagamihara, Kanagawa 229-8510, Japan}
\altaffiltext{3}{Max-Planck-Institut f\"ur Extraterrestrische Physik, Giessenbachstra{\ss}e, 85748 Garching, Germany}
\altaffiltext{4}{Department of Astronomy, Kyoto University, Kyoto 606-8502, Japan}



\begin{abstract}
We have carried out a precise energy spectral analysis of the
super-luminal jet source GRS1915+105 observed with ASCA six times from
1994 to 1999.  The source was so bright that most SIS data suffered
from event pileup.  We have developed a new technique to circumvent
the pileup effect, which enabled us to study the spectrum in detail
and at high resolution ($\Delta E/E \approx$ 2 \%).  In the energy
spectra of 1994 and 1995, resonant absorption lines of Ca {\sc xx}
K$\alpha$, Fe {\sc xxv} K$\alpha$, Fe {\sc xxvi} K$\alpha$, as well as
blends of the absorption lines of Ni {\sc xxvii} K$\alpha$ + Fe {\sc
xxv} K$\beta$ and Ni {\sc xxviii} K$\alpha$ + Fe {\sc xxvi} K$\beta$,
were observed.  Such absorption lines have not been found in other
objects, except for iron absorptions lines from GRO J1655$-$40,
another super-luminal jet source (Ueda et al.\ 1998).  We carried out a
``curve of growth'' analysis for the absorption lines, and estimated
column densities of the absorbing ions.  We found that a plasma of
moderate temperature (0.1--10 keV) and cosmic abundance cannot account
for the observed large equivalent widths.  The hydrogen column density
of such plasma would be so high that the optical depth of Thomson
scattering would be too thick ($N_H\gtrsim10^{24}$cm$^{-2}$).  We
require either a very high kinetic temperature of the ions ($\gtrsim
100$ keV) or extreme over-abundances ($\gtrsim 100$ $Z_\odot$).  In
the former case, the ion column densities have reasonable values of
$10^{17} - 10^{18}$ cm$^{-2}$.  We modeled the absorber as a
photo-ionized disk which envelops the central X-ray source.  Using a
photo-ionization calculation code, we constrain physical parameters of
the plasma disk, such as the ionization parameter, radius, and
density.  Estimated parameters were found to be consistent with those
of a radiation-driven disk wind.  These absorption-line features may
be peculiar to super-luminal jet sources and related to the jet
formation mechanism.  Alternatively, they may be common
characteristics of super-critical edge-on systems.
\end{abstract}


\section{Introduction}
GRS 1915+105 was discovered as a transient X-ray source with
WATCH/Granat (Castro-Tirado, Brandt, \& Lund 1992), and later
recognized as a super-luminal Galactic jet source at a distance of
12.5 kpc (Mirabel \& Rodr\'{\i}guez 1994).  The velocity of the jet
and its inclination were estimated by Mirabel \& Rodr\'{\i}guez (1996)
to be 0.92 $c$ and 70$^\circ$, respectively .  The optical type of the
companion star and binary parameters are as yet unknown.  The
formation of the relativistic jet is considered to be related to
instabilities in a super-critical accretion disk (Belloni et al.\
1997; Mirabel et al.\ 1998), but the precise formation mechanism is
not yet clear.  GRS 1915+105 is extremely important because it is a
potential key to understanding the production of jets, which is
relevant to astrophysical systems ranging from binaries to AGNs.  To
obtain information on the environment of the central engine of jet
sources, X-ray spectroscopy with a fine resolution is desirable.

GRS 1915+105 was observed with the X-ray astronomy
satellite ASCA, and surprisingly, absorption features of calcium and
iron K structure were found in the spectrum (Ebisawa 1997a; Ebisawa
1997b; Kotani et al.\ 1997a; Ebisawa et al.\ 1998).  These features were
interpreted as absorption lines due to K$\alpha$-resonant scattering
of helium-like and hydrogen-like ions (Kotani et al.\ 1999a; Kotani et
al.\ 1999b).  Such X-ray absorption lines have never been found in
other X-ray binaries except for GRO J1655$-$40, another super-luminal
jet source (Ueda et al.\ 1997; Ueda et al.\ 1998).  Ueda et al.\
(1997; 1998) identified helium-like and hydrogen-like
iron-K$\alpha$ absorption lines, and estimated the physical condition
of the plasma scattering iron-K$\alpha$ photons.  Since two of the
super-luminal jet sources show absorption lines, it is natural to
consider the lines in the context of jet production.  The absorption lines
may be a byproduct of the unknown jet formation mechanism.  In this
paper, we discuss the identification and origin of the X-ray
absorption lines in GRS 1915+105 in detail.

\section{Observations}
GRS 1915+105 was observed with ASCA six times from 1994 to 1999
(Table~\ref{kotani:tbl:obs}).  In this paper we mainly report and
discuss the first two observations in 1994 and 1995, in which
characteristic absorption lines were detected.  In 1994, the source
was observed as a Target-Of-Opportunity (TOO), responding to the
report of a hard X-ray outburst (Sazonov et al.\ 1994).  The SIS was
always operated in 1-CCD BRIGHT mode except for the observation in
1995, when the SIS was in 1-CCD FAINT mode.  The telemetry limit is
256 counts/s/SIS in BRIGHT mode at the highest telemetry rate, and 64
counts/s/SIS in FAINT mode.  Therefore, the telemetry was saturated
and a significant fraction of SIS events was lost in the 1995
observation.  The GIS was always operated in PH mode with the nominal
bit assignment (Ohashi et al.\ 1996). The source flux in the 2--10 keV
band was 0.4, 0.8 and 0.9 Crab for the observations in 1994 (Nagase et
al.\ 1994), 1995 and 1996, respectively. Due to the high counting
rate, the GIS spectra suffered from the both dead-time effects and
telemetry saturation, which we correct for in the present analysis
(Makishima et al.\ 1996).

\placetable{kotani:tbl:obs}

\section{Data Reduction}
\subsection{Extracting SIS and GIS Spectra}
Event pileup effects must be removed from SIS data to perform precise
spectroscopy for bright sources like GRS1915+105.  We have developed a
new method which can correct both of the two symptoms of event pileup,
namely, the reduction in counting rate and hardening of the spectrum.
The correction method we applied is described in detail in
Appendix~A\@.  The spectra taken in 1994 and 1996 were corrected with
this code.  As for the 1995 data, pileup correction is not required,
because a significant fraction of the image including the part where
pileup effects would be most severe was lost due to the telemetry
saturation.  The response function of the X-ray telescope (XRT)
(Serlemitsos et al.\ 1996) for each selected region was calculated.
Background subtraction was not performed since the source was brighter
than the background by two orders of magnitude over the entire energy
range.  The gains of SIS-0 and SIS-1 were calibrated based on the
instrumental gold-M-edge and silicon-K-edge features.  The two spectra
were then combined and a systematic error of 2 \% was added to each
spectral bin, in quadrature with the statistical error.

GIS events were collected within a circular region with a radius of
$6'$ centered on the source position for each sensor, and the
corresponding XRT response functions were calculated. Background was
again not subtracted.  After gain calibration using the instrumental
gold M edge, the GIS-2 and GIS-3 spectra were combined.  A systematic
error of 2 \% was included for each energy bin of the spectra.

The resultant SIS and GIS spectra were found to be not completely
consistent with each other: The SIS spectra were systematically harder
than the GIS spectra even after the pileup correction was applied.  We
also tried other pileup correction codes (e.g., Ebisawa et al.\ 1996;
Ueda et al.\ 1997), but the results were essentially the same.  Here
we consider the GIS spectrum to be more reliable than the SIS to study
the continuum spectral shape, since the SIS data for bright sources
have calibration uncertainties other than pileup, e.g., uncorrectable
DFE and the echo effect (Yamashita et al.\ 1997 and references
therein).  On the other hand, the gain dependence on count rate of the
GIS is well understood (Makishima et al.\ 1996).  We do not pursue the
GIS and SIS spectral difference further in this paper, but rather
discuss the spectral features common to both the SIS and GIS\@.  Note
that pileup is a global effect, so it will hardly affect any {\em
sharp}\/ spectral features like the absorption lines which we are
mostly interested.

\section{Spectral Analysis Results}
\subsection{Continuum Spectrum}
We tried several spectral  models, and found that 
both the SIS and GIS spectra are roughly expressed with a
partially attenuated cut-off-power-law model:
$$
\exp[-\sigma(E)\,N_{\rm H1}] \times \{f\,\exp[-\sigma(E)\,N_{\rm H2}] +
(1-f)\}\times A\, \exp[-E/E_{\rm fold}]\,E^{-\Gamma}.
$$
Where $E$ is the incident photon energy, $N_{\rm H1}$ and $N_{\rm H2}$
are the attenuating hydrogen column densities, $f$ is the partial
covering factor, $A$ is the normalization factor, $E_{\rm fold}$ is
the folding energy, $\Gamma$ is the photon index, and $\sigma(E)$ is
the cross section of the neutral solar-abundance gas. It should be
stressed that we adopt this model as a simple mathematical expression
of the continuum spectra, and do not imply that it is the physical
emission mechanism of the source.  For example, it is possible that
the continuum spectrum consists of more complicated components.  The
origin of the continuum emission will be discussed in forthcoming
papers.

\begin{figure}
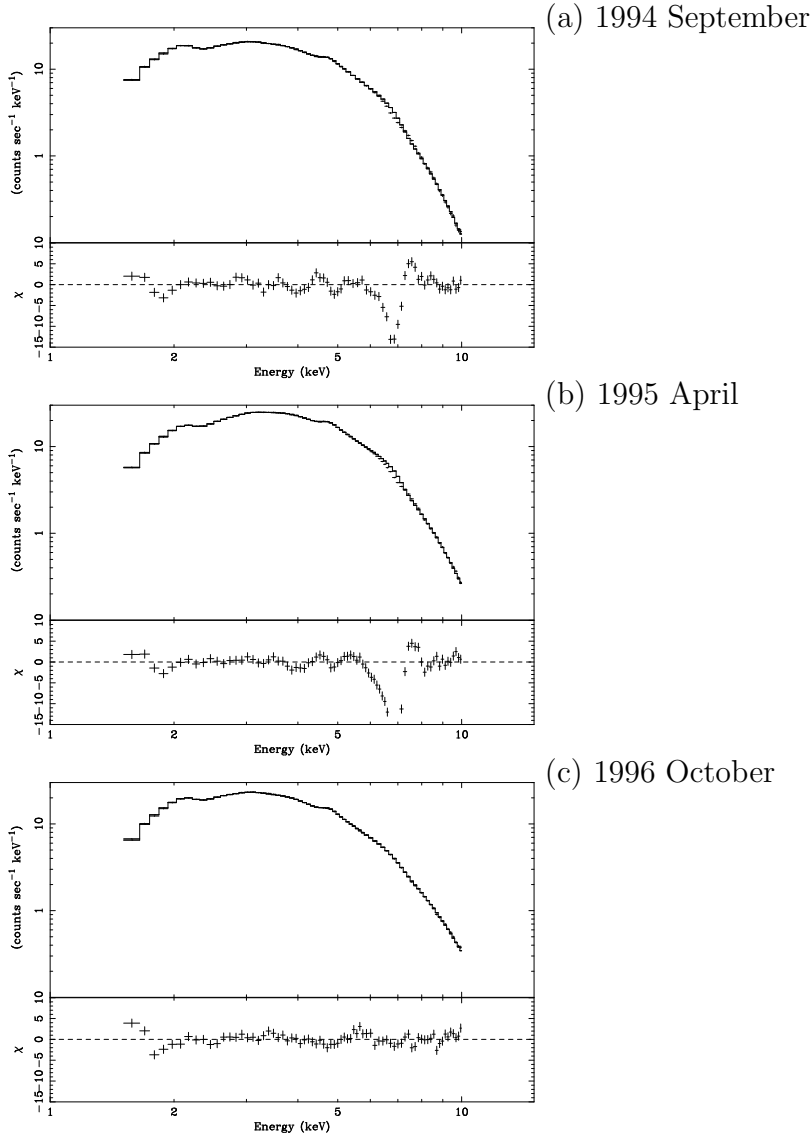

\resizebox{7cm}{!}{\rotatebox{270}{\includegraphics{spec94.g.ps}}}
(a) 1994 September

\resizebox{7cm}{!}{\rotatebox{270}{\includegraphics{spec95.g.ps}}}
(b) 1995 April

\resizebox{7cm}{!}{\rotatebox{270}{\includegraphics{spec96.g.ps}}}
(c) 1996 October

\caption{GIS Spectra.  (a) 1994 September, (b) 1995 April, and (c)
1996 October.  Observed data (cross), best-fit model (histogram), and
residuals (bottom panel) are plotted.  Each error bar includes
systematic uncertainties of 2 \%.  The data of GIS-2 and GIS-3 were
combined.  Dead time and telemetry saturation were not corrected.  The
energy range 6--8 keV in the data of 1994 and 1995 was excluded in
fitting.  The structure at 4.7 keV is due to an instrumental Xe-L
edge.
\label{kotani:fig:specgis}}
\end{figure}

We found  that there is a complicated feature near 6 keV in the
spectra of 1994 and 1995, which is not well accommodated by the
continuum model alone.  To fit the GIS continuum spectra in 1994 and 1995 satisfactorily, 
we had to exclude the energy range  6--8 keV\@.
In the fit of the spectrum in  1996, the range 6--8 keV was included
because no complicated feature was seen in there.
The GIS spectra and the best-fit parameters are shown in Fig.~\ref{kotani:fig:specgis} and
Table~\ref{kotani:tbl:bestfitgis}, respectively. 
As clearly seen,  deep and broad absorption features exist at
6.5--7 keV in the residuals of 1994 and 1995, as well as another absorption
feature at $\sim$ 8 keV\@.   In addition, there is a
dip feature at 4 keV\@. In 1996, these absorption and dip
features disappear or become shallower.

\placefigure{kotani:fig:specgis}

\placetable{kotani:tbl:bestfitgis}

\begin{figure}

\resizebox{7cm}{!}{\rotatebox{270}{\includegraphics{spec94.s.ps}}}
(a) 1994 September

\resizebox{7cm}{!}{\rotatebox{270}{\includegraphics{spec95.s.ps}}}
(b) 1995 April

\resizebox{7cm}{!}{\rotatebox{270}{\includegraphics{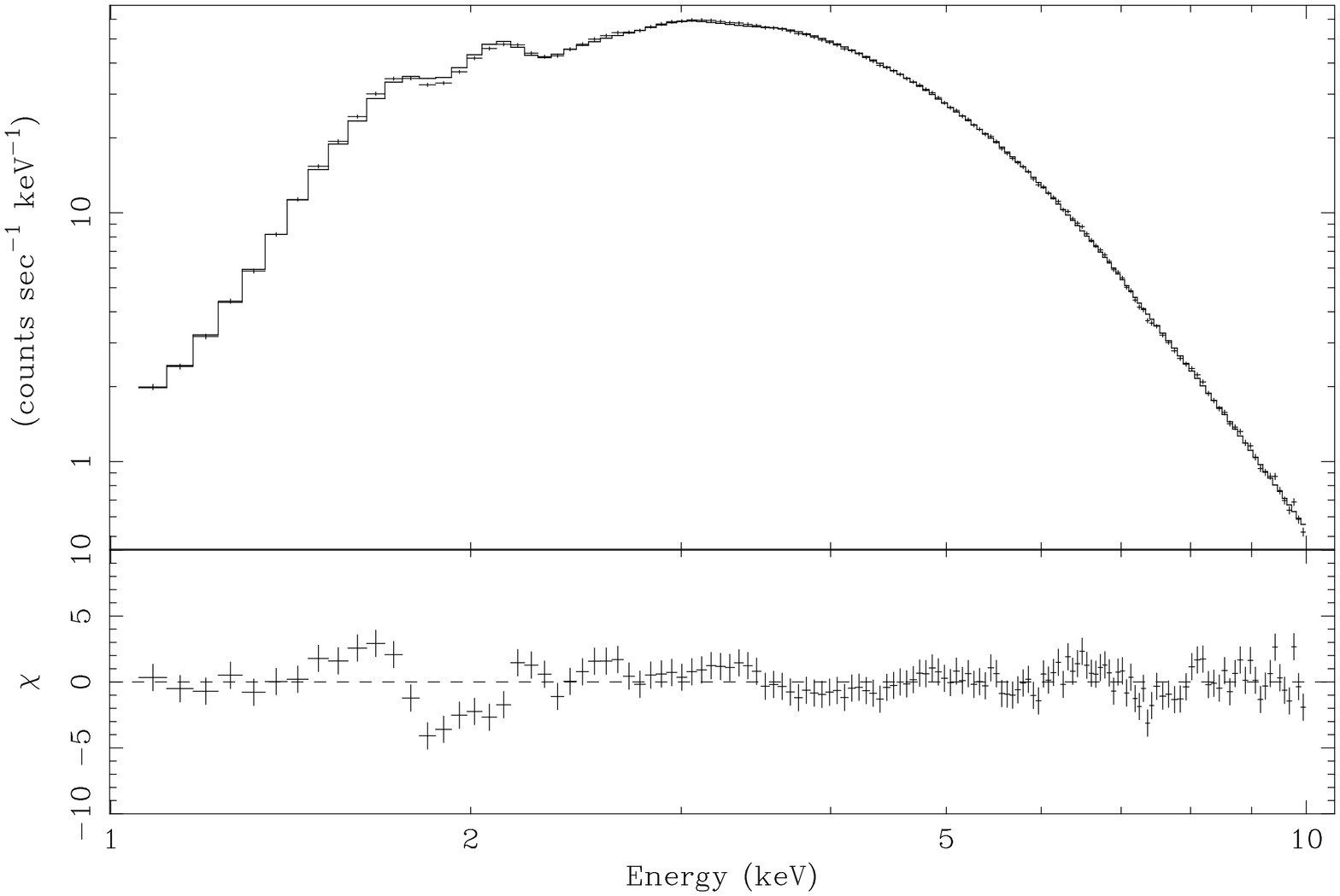}}}
(c) 1996 October

\caption{SIS Spectra.  (a) 1994 September, (b) 1995 April, and (c)
1996 October.  Observed data (cross), best-fit model (histogram),
residuals (crosses in lower panel), and residuals for the model without
absorption lines (solid line in lower panel) are plotted.  Absorption
features are clearly seen in the solid curve.  As for the data of 1994
and 1995, only the range around the iron absorption features are
plotted, while the overall spectrum is plotted for 1996.  The data of
SIS-0 and SIS-1 were combined.  The structure around 2--2.5 keV is
instrumental.
\label{kotani:fig:specsis}}
\end{figure}

To investigate these features, we use SIS spectra, which have a
superior energy resolution.  The SIS spectra were fitted with the
tentative continuum model determined from the GIS\@.  Complicated
absorption features were clearly seen in the spectra of 1994 and 1995
in Fig.~\ref{kotani:fig:specsis}.  In particular, two narrow
absorption lines (below and above 6.8 keV) are recognized in 1994.

\placefigure{kotani:fig:specsis}

\subsection{Absorption Line Features}
Next, we tried to fit the spectral features at the iron K energy band
for the 1994 and 1995 SIS data.  The 5--10 keV range was initially
fitted with a cut-off-power-law model, and Gaussians with negative
normalizations were added one by one.  Intrinsic Gaussian widths were
allowed to be free but tied together.  Four negative Gaussians were
added to the models of 1994 and 1995, respectively, to achieve
$\chi^2/\nu=80.5/56$ and 112.3/58, respectively, which are not
satisfactory yet.  We found that inclusion of an emission line at 6.4
keV to the model of 1994 and an absorption edge at 7.117 keV to that
of 1995 improves the fits to an acceptable level.  The best-fit
parameters are shown in Table~\ref{kotani:tbl:bestfitsis}.  The
negative Gaussians are tentatively identified in
Table~\ref{kotani:tbl:bestfitsis} and referred to hereafter as
negative Gaussians A, B, and so on.  Adding an absorption edge to the
model of 1994 or an emission line to that of 1995 does not improve the
fits.

We also tried absorption edges instead of absorption lines, so that
each negative Gaussian in Table~\ref{kotani:tbl:bestfitsis} is
replaced with an absorption edge.  We found all the negative Gaussians
cannot be replaced simultaneously, and at least one or two negative
Gaussians are required to fit the spectra.  These results suggest that
absorption lines do in fact exist in the spectra of 1994 and 1995.

To check the consistency between the SIS and GIS, both data were
fitted simultaneously.  The equivalent widths and center energies of
the emission and absorption lines, and optical depths of the edges
were set equal for both data, while the parameters of the continuum
model were allowed to be free.  As a result, the best-fit parameters
for the absorption features were consistent with those obtained from
the SIS fit alone.  Therefore, the absorption line parameters we have
obtained are confirmed by both SIS and GIS, and we consider their
detection to be unrelated to SIS pileup effects.

\placetable{kotani:tbl:bestfitsis}

To measure the absorption feature around the calcium K energy ($\sim 4$
keV), the spectral continuum between 2.5 keV and 6 was fitted with an
attenuated-power-law model.  Then an absorption-line model, negative
Gaussian, was added to the fit model.  It was found that the
absorption feature of 1994 is fitted well with a negative Gaussian,
while inclusion of a negative Gaussian does not improve the fit of
1995.  The best-fit parameters are shown in
Table~\ref{kotani:tbl:bestfitsisca}.

\placetable{kotani:tbl:bestfitsisca}

\section{Discussion}
\subsection{Interpretation of the Absorption Lines}
We notice the center energies of the negative Gaussians shown in
Table~\ref{kotani:tbl:bestfitsis} are consistent with or very close to
the K-line energies of iron or nickel ions.  Candidate resonant and
inter-combination lines are listed in
Table~\ref{kotani:tbl:candidates} (forbidden lines are not included).
The energies of the negative Gaussians A and B coincide with those of
Fe {\sc xxv} K$\alpha$ ($1s^12p^1$ $^1{\rm P}_1$ and $1s^12p^1$
$\rm^3P_{1,2}$ in Table~\ref{kotani:tbl:candidates}) and Fe {\sc xxvi}
K$\alpha$ ($2p^1$ $\rm^2P_{1/2,3/2}$), respectively.  The negative
Gaussian C is considered to be Fe {\sc xxv} K$\beta$ ($1s^12p^1$
$\rm^1P_1$) or Ni {\sc xxvii} K$\alpha$ ($1s^12p^1$ $\rm^1P_1$ and
$1s^12p^1$ $\rm^3P_{1,2}$), although nickel is less abundant than iron
in a plasma of cosmic abundance.  The negative Gaussian D resides just
between the Ni {\sc xxviii} K$\alpha$ ($2p^1$ $\rm^2P_{1/2,3/2}$) and
Fe {\sc xxvi} K$\beta$ ($3p^1$ $\rm^2P_{1/2}$) energies.  These
coincidences strongly support the interpretation that the absorption
features consist of iron or nickel resonant absorption lines.
Hereafter, we discuss the physics of the system assuming that the
negative Gaussians A, B, C, and D are absorption lines of Fe {\sc xxv}
K$\alpha$, Fe {\sc xxvi} K$\alpha$, Fe {\sc xxv} K$\beta$ + Ni {\sc
xxvii} K$\alpha$, and Fe {\sc xxvi} K$\beta$ + Ni {\sc xxviii}
K$\alpha$, respectively,

\placetable{kotani:tbl:candidates}

\subsection{Curve of Growth Analysis}
The observed spectral features are obviously different from the
P-Cygni profile which accompanies strong emission lines, and thus the
absorption is probably not caused by a spherical wind.  As for the
origin of the absorption lines, we will consider line-scattering
material anisotropically irradiated or anisotropically distributed
around the central source.  To estimate amount of the line-scattering
matter in the line of sight, we calculate the ``curve of growth'',
namely, the equivalent widths of an absorption line as function of the
column density of scattering ions.

We summarize here the basic equations we used to calculate the curves of
growth.  Let us assume particles with the column density $N$ producing
an absorption line at frequency $\nu_0$ by resonant scattering.
The optical depth $\tau(\nu)$ is expressed as
\begin{equation}
\tau(\nu) = N s \phi_a(\nu),
\end{equation}
and the normalized equivalent width
$W_\nu/\nu_0$ is
\begin{equation}\label{ew}
        \frac{W_\nu}{\nu_0} = \frac 1 {\nu_0} \int (1-\exp[-\tau(\nu)])d\nu,
\end{equation}
respectively, where $s \phi_a(\nu)$ is the
averaged cross section of scattering particles, normalized as $\int
\phi_a(\nu) d\nu =1$  (see Spitzer 1978, section 3.4).
In CGS unit, the transition probability $s$ is expressed as
\begin{equation}
s = \frac{\pi f_{lu}e^2}{m_{\rm e}c}(1-\frac{b_u}{b_l}\exp[-h
\nu_0/kT]),
\end{equation}\label{kotani:eq:transition}
where $f_{lu}$ is the oscillator strength, and the subscripts $u$ and
$l$ denote values in the upper and lower states, respectively.  The
factor $(b_u/b_l)\exp[-h \nu_0/kT]$, expressing the effect of induced
emission, is defined to be equal to the actual rate of the density of
ions in the upper state to that in the lower (Spitzer 1978, section
2.4).  We assumed that the density of the ions in the upper state is
negligible and omit the factor from the following calculation.  The
validity of the assumption is checked in
\S~\ref{kotani:subsec:validity}.
If scattering particles obey the Maxwell-Boltzmann distribution with
temperature $T$,
$\phi_a(\nu)$ is written as a Voigt function $H(a, u)$ with
\begin{eqnarray}
\phi_a(\nu) &=  &\frac c {\sqrt\pi\nu_0b} H(a,u)\\
H(a,u) &= &\frac a \pi \int \frac{\exp[-y^2]dy}{a^2+(u-y)^2}\\
a &= &\frac {c}{4\pi\nu_0b}\sum_{l}A_{ul}\\
u &= &\frac c {\nu_0b}(\nu-\nu_0)\\
b &= &\sqrt\frac{2kT}{m\rm_i},
\end{eqnarray}
where $A_{ul}$ are the Einstein coefficients related to the oscillator
strengths as
\begin{equation}
A_{ul} = \frac{8\pi^2\nu_0^2e^2}{m_{\rm e}c^3}
\frac {g_l}{g_u}f_{lu}
\end{equation}
(see Rybicki \& Lightman 1979, section 10.6).  Even if the energy
distribution of particles is dominated by turbulence or other bulk
motion rather than thermal motion, these equations are
still applicable as a good approximation.  In that case, $T$
represents the velocity dispersion within the scattering gas.  With
these formulas and constants in Table~\ref{kotani:tbl:candidates}, the
equivalent widths are calculated according to equation (\ref{ew}), and
the curves of growth are plotted in Fig.~\ref{kotani:fig:curve}.

Using the curves of growth thus calculated, the column densities of the
ions $N_{\rm ion}$ responsible for the observed absorption lines may
be obtained.  Since Fe {\sc xxv} K$\alpha$ and Fe {\sc xxvi} K$\alpha$
lines are considered not contaminated by other ion species, $N_{\rm
Fe\, XXV}$ and $N_{\rm Fe \, XXVI}$ are directly derived from the
observed equivalent widths, assuming a temperature.  From these column
densities, equivalent widths of Fe {\sc xxv} K$\beta$ and Fe {\sc
xxvi} K$\beta$ are estimated using the curves of growth.  Then, these
estimated equivalent widths are subtracted from the observed
equivalent widths of the blends of Ni {\sc xxvii} K$\alpha$ + Fe {\sc
xxv} K$\beta$, or of Ni {\sc xxviii} K$\alpha$ + Fe {\sc xxvi}
K$\beta$, to obtain the equivalent widths of Ni {\sc xxvii} K$\alpha$
or Ni {\sc xxviii} K$\alpha$.  Column densities of $N_{\rm Ni\, XXVII}$
and $N_{\rm Ni\, XXVIII}$ are obtained using the curves of growth
again.  Ion column densities thus derived are shown in
Table~\ref{kotani:tbl:column} for two extreme temperatures, 1000 keV
and 0.1 keV.

If the temperature is as high as 1000 keV (high-temperature limit),
all the observed values of the equivalent width will be found in the
``linear part'' of the curves of growth, while if temperature is as
low as 0.1 keV (low-temperature limit), they will be in the
``square-root section'' of the curves.  It also should be noted that
the column densities shown in the Table~\ref{kotani:tbl:column} are lower
limits for both temperatures, since line photons scattered from
matter located out of the line of sight would reduce the absorption
line equivalent widths.  That effect is not taken into account here.

The derived column densities for the low-temperature limit are larger
than those for the high-temperature limit by several orders of
magnitude.  If the column densities are so large as the
low-temperature calculation suggests, and if abundances of the
absorbing plasma is not very different from the solar value, the
optical depth of the plasma for the Thomson scattering will exceed
unity.  This is unlikely, since the modification of the absorption
line features that would be present due to Thomson scattering is not.
It is plausible that either iron and nickel are significantly
over-abundant, or that the kinematic temperature is as high as the
high-temperature limit suggests.  However, at such high temperature,
iron and nickel would be fully ionized and absorption lines should not
be observed.  It may be that the ionization temperature of the
absorbing plasma is lower than the kinematic temperature, or that the
absorbing matter consists of several parts with different bulk
velocities whose dispersion is comparable to thermal velocity of iron
atoms at $kT>$ 100 keV\@.  In following discussion, we first consider
the high-temperature limit in Table~\ref{kotani:tbl:column}, which
presumably gives the lower limit of column density.

\placefigure{kotani:fig:curve}

\placetable{kotani:tbl:column}

\subsection{Photo-ionized Plasma Model}
We note that the spectral parameters changed between the observations
of 1994 and 1995.  The luminosity in the 1995 observation was higher
than in 1994, and the ratio $N_{\rm Fe\, XXVI}/N_{\rm Fe \, XXV}$
became larger, suggesting that the absorbing matter was more ionized.
This behavior is consistent with characteristics of a photo-ionized
plasma.  Thus, we assume that the source of absorption lines in 1994
and 1995 is a photo-ionized plasma irradiated by the central source.
In 1996, the system was considered to be in a different state,
because, not only were absorption lines absent, but also the photon
index and folding energy of the spectral continuum were different.  We
calculated electron temperature and ion-population distribution of the
photo-ionized plasma with XSTAR (Kallman \& Krolik 1996), and searched
for parameters which are consistent with those in the high-temperature
limit in Table~\ref{kotani:tbl:column}.  To simplify the calculation,
the plasma geometry was assumed to be isotropic around the source.  It
should be noted that assumption of isotropy is adopted only in the
calculation, and that the realistic geometry of the line-absorbing
plasma must be anisotropic, since otherwise absorption lines would be
canceled by emission lines from out of the line of sight.  In the
calculations, abundance was assumed to be solar.  The
absorption-corrected 2--10 keV luminosity was estimated from
Table~\ref{kotani:tbl:bestfitgis} assuming a distance of 12.5 kpc, and
used as total luminosity of the model.  Continuum spectral parameters
were drawn from  Table~\ref{kotani:tbl:bestfitgis}.  We arbitrarily fixed the ionization
parameter $\xi = L/(nr^2)$ at the innermost boundary of the plasma to
be $10^{5.5}$.  Our results depend only weakly on this
value, because iron and nickel is fully ionized with such a high ionization parameter at the innermost
boundary, and thus column
density of hydrogen-like or helium-like ions is not sensitive to the
innermost condition.  The assumed geometry is schematically shown in
Fig.~\ref{kotani:fig:geometry}.  We searched a combination of plasma
density and outer radius to give $N_{\rm Fe \, XXVI}$ and $N_{\rm Fe
\, XXV}$ consistent to observation.  The determined parameters are
shown in Table~\ref{kotani:tbl:photo}, and expected column density of
ions are shown in Table~\ref{kotani:tbl:calccolumn}.  Obviously, the
expected column density of nickel ions in
Table~\ref{kotani:tbl:calccolumn} is much lower than the observation
by one or two orders of magnitude.  Therefore, we conclude that the
relative abundance of nickel to iron is larger than that of solar
value.

We next searched for a combination of plasma parameters to reproduce
the column density in the low-temperature limit in
Table~\ref{kotani:tbl:column}.  Since the metal abundance was found to
be non-solar, there was no reason to fix them to the solar value.  We
found that if both the iron and nickel abundances are multiplied by
$\sim 100$, the resultant photo-ionized plasma gives the column
densities in the low-temperature limit.  Other parameters, such as the
electron density, the ionization parameter at the outer boundary, and
the outer radius of the plasma were hardly affected.  Although the
column densities are different between the high-temperature-limit and
low-temperature-limit cases, the column-density ratios of Fe {\sc xxv}
to Fe {\sc xxvi} are not much different between the two cases, and
thus both yield similar values of $\xi_1$.  (Subscript``0'' and ``1''
denote values at the innermost and outermost boundary, respectively.)
Thus we consider that $\xi_1$ and $r_1$ in
Table~\ref{kotani:tbl:photo} are reliable, regardless of the kinematic
temperature, the metal abundances, the electron density, or the
innermost radius of the plasma.

\placefigure{kotani:fig:geometry}

\placetable{kotani:tbl:photo}

\placetable{kotani:tbl:calccolumn}

\subsection{Geometry}
To produce absorption lines, the solid angle of the line-scattering
material from the source must be small, otherwise the absorption lines
would be offset by an emission line counterpart.  From a moderate
assumption that the solid angle is less than $4\pi\exp[-1]$, the disk
half thickness, $h$, at the radius $r_1$ is constrained as $h/r_1
\lesssim 0.4$, where $h/r_1$ is related to the solid angle
as $\Omega = 1/\sqrt{1+(h/r_1)^2}$.  Therefore, the half opening angle
of the plasma, $\theta$, is constrained to be $\theta < 20^\circ$.
The inclination of the jets is $70^\circ$ to our line of sight
(Mirabel \& Rodor\'{\i}guez 1993).  We might reasonably assume that the
immediate neighborhood of the central source (e.g., the accretion
disk) is also so inclined. From these two constraints, we conclude that
the half thickness of the reprocessing plasma is $\sim 0.4 r_1$.
Thus, total mass of the reprocessing plasma is estimated as $\sim 4.4
\times 10^{-11}$ M$_\odot$ (1994) and $\sim 1.5 \times 10^{-10}$
M$_\odot$ (1995).

Note that $n\rm_p$ in Table~\ref{kotani:tbl:photo} is dependent on
assumed geometry and has a large uncertainty, while $\xi_1$ and $r_1$
are rather model independent and thus reliable.  For example, a
toroidal plasma distribution around the source can also explain the
observed properties, and in this case the path in the plasma along the
line of sight will be shorter.  Therefore the proton density $n\rm_p$
will be larger, conserving column density $(r_1-r_0) \times n_{\rm
p}$.  The other parameters $\xi_1$ and $r_1$ would not be changed
much, because the former is determined from the ratio $N_{\rm Fe\,
XXVI}/N_{\rm Fe\, XXV}$, and the latter is determined from the
relation $\xi_1 = L/(N_{\rm e}r_1^2)$.

\subsection{Nature of the Reprocessing Plasma}
From the discussion above, we can construct a picture of the plasma
producing the absorption lines.  It is located at $\sim 10^{11}$ cm
from the source, constrained to be within $\theta<20^\circ$ of the
disk plane, and has a total mass of $10^{-11}$ to $10^{-10}$
M$_\odot$.  The kinetic temperature of iron ion is either high or low.
If the metal abundance is comparable to the solar value, then the
kinematic temperature must be as high as or higher than 100 keV, which
corresponds to a velocity of $\sim10^8$ cm s$^{-1}$.  If the
reprocessing plasma is hotter than 100 keV, it must be observed within
10 s of its production to avoid full ionization of iron and nickel.  A
post-shock flow within 10 s after the shock is a candidate for such a
hot plasma.  Both the thermal motion of each ion and the bulk motion
of the plasma may give such a high kinematic temperature.  An inward
or outward bulk flow with a velocity gradient or randomly moving blobs
may be the X-ray reprocessor.

As a candidate of such a flow, we suggest a radiation-driven disk
wind.  If gas at rest exists in the vicinity of the super-Eddington
X-ray source, it would be blown out by Thomson scattering.  The
terminal velocity $v_{\rm wind}$ would be
\begin{equation}
v_{\rm wind} \sim \sqrt{ \frac {L\sigma_{\rm Th}}{4\pi rcm_{\rm p}}},
\end{equation}
where $L$ is the luminosity of the source, $\sigma_{\rm Th}$ is the
cross section of Thomson scattering, $r$ is the initial distance of
the gas from the source, and $m\rm_p$ is the mass of a proton.  If
such a gas is supplied from the accretion disk, we would observe a
stream with a radial-velocity gradient from 0 to $v_{\rm wind}$.
Substituting $r = 10^{11}$ cm and $L = 4\times 10^{38}$ erg s$^{-1}$,
a terminal velocity of $v_{\rm wind} = 10^8$ cm s$^{-1}$ is obtained,
which coincides with the velocity inferred in the curve-of-growth
discussion.  It should be noted that most of the centroid energies of
the absorption lines in Table~3 are blue-shifted from those in
Table~5, which would be expected from the absorption lines from an
outward flow with a velocity of $10^8$ cm s$^{-1}$.  Thus a
radiation-driven disk wind at $ \sim 10^{11}$ cm can well explain the
observed absorption line features.  Because the disk wind would be
refreshed in every $\sim 10^3$ s, the outflow rate is estimated as
$10^{-7} - 10^{-6}$ M$_\odot$ yr$^{-1}$.

On the other hand, if the plasma is very metal rich, the assumption of
high-temperature is no longer necessary.  The high column density of
ions may be explained by low-temperature, metal rich plasma, if the
metal abundance is $100$ times higher than the solar value.  Even in
that case, the conclusion on the geometry of the plasma would not be
changed appreciably.  The nickel abundance of the jet material of
SS~433 has been found to be a few tens times higher than the solar
value (Kotani et al.\ 1997b), which suggests that anomalous abundances
may not be rare in binary systems with strong jets.  Thus they may be
possible that iron and nickel are abundant in the environment of
GRS~1915+105, and that the reprocessing plasma has a velocity
dispersion smaller than $10^8$ cm s$^{-1}$ or a lower temperature than
100 keV\@.

\subsection{Alternative situations}\label{kotani:subsec:validity}
In the analysis above, the term of the induced emission
$(b_u/b_l)\exp[-h \nu_0/kT]$ in equation~(\ref{kotani:eq:transition})
was neglected.  The validity of the assumption can be confirmed as
follows.  Suppose that a reprocessing plasma is illuminated by an external
photon source.  An ion in the plasma receives photons with a mean
interval of $1/(\sigma_{\rm res}F(\nu_0))$, and emits each photon
after a time $\sim 1/A_{ul}$, where $F(\nu_0)$ is the photon flux of
the source and $\sigma_{\rm res}$ is the resonant-scattering cross
section.  Therefore the ratio of the number density of ions at upper
state to those at lower state is $\sigma_{\rm res}F(\nu_0)/A_{ul}$.
This is an overestimation because photons at $\nu_0$ are reflected by
the plasma surface due to resonant scattering.  If the ratio is much
less than unity, the term of induced emission is negligible.
Substituting $\sigma_{\rm res} = c^2/(8\pi\nu_0^2)$, the ratio is
estimated as
\begin{equation}
\frac {b_u}{b_l}\exp[-h\nu_0/kT] 
	\sim \frac {c^2F(\nu_0)}{8\pi\nu_0^2A_{ul}} 
	\sim 10^{-10} 
	\left ( 
		\frac L {10^{39}\rm \; erg\; s^{-1}}
	\right )
	\left ( 
		\frac R {10^{11}\rm \; cm}
	\right )^{-2},\label{kotani:eq:ratio}
\end{equation}
which is much less than unity.  Thus induced emission is shown to be
negligible in the case of the photo-ionized plasma treated here.  

There is another case where the induced-emission term is important.
The ratio~\label{kotani:eq:ratio} approaches unity on the surface of a
neutron star ($R = 10^6$ cm).  Under the presence of the induced
emission, more ions are necessary to produce an absorption line.  As
an estimation of column density of Fe {\sc xxvi}, we adopt $10^{20.4}$
cm$^{-2}$ from Table~6.  The corresponding hydrogen column density
would be $10^{25}$ cm$^{-2}$, assuming solar abundance.  To avoid full
ionization of iron, the ionization parameter must be $\xi \leq
10^{4}$, and thus the density of the plasma must be $\geq 10^{22}$
cm$^{-3}$, and the height of the atmosphere must be
$\stackrel{<}{\sim}10^{3}$ cm.  Although the scale height of the
atmosphere at $kT \sim 100$ is larger than $10^{3}$ cm by an order of
magnitude, such an atmosphere may produce absorption lines.  However,
the absorption lines made by the atmosphere would be gravitationally
red-shifted, which was not observed.  The atmosphere of a neutron star
is unlikely to be the origin of the absorption lines.  Thus we
conclude that induced emission can be neglected in the usual
situations where equation~(\ref{kotani:eq:ratio}) is applicable.

\section{Summary}
Absorption lines of Ca {\sc xx} K$\alpha$, Fe {\sc xxv} K$\alpha$, Fe
{\sc xxvi} K$\alpha$, a blend of Ni {\sc xxvii} K$\alpha$ + Fe {\sc
xxv} K$\beta$, and a blend of Ni {\sc xxviii} K$\alpha$ + Fe {\sc
xxvi} K$\beta$ have been discovered in the energy spectra of the
transient jet source GRS 1915+105.  These features can be explained by
resonant scattering in a disk-like photo-ionized plasma.  From the
``curve-of-growth'' analysis, column densities of the ions were
determined as in Table~\ref{kotani:tbl:column}.  Adopting the values
of high-temperature limit, the physical parameters of the plasma were
determined to be:
\begin{itemize}
\item outer radius $r_1 \sim 10^{11}$ cm
\item half thickness $h = 0.4 r_1$
\item density $n_{\rm p} \sim 10^{11}$ cm$^{-3}$
\item ionization parameter at the outer boundary of the plasma $\xi = 10^{3.8}$
\item kinematic temperature of scattering ions $> 100$ keV, i.e., $>10^8$
cm s$^{-1}$, or metal
abundance is as large as 100 times of the solar value.
\end{itemize}
The kinematic temperature of the scattering ions should be very high
to account for the observed large equivalent widths of the absorption
lines, if solar abundance is assumed.  If the temperature is so high,
the plasma may not be in equilibrium, with an ionization temperature
lower than the kinematic temperature.  Thus the high kinematic
temperature may be explained by a bulk motion, such that a velocity
difference up to $\sim 10^8$ cm s$^{-1}$ within the plasma flow, or
perhaps plasma blobs moving randomly.  We found that a
radiation-driven accretion flow with a mass outflow rate of $10^{-7} -
10^{-6}$ M$_\odot$ yr$^{-1}$ can account for the absorption lines.
Alternatively, if the plasma is extremely metal rich, such a high
kinematic temperature is not necessary.  Regardless of the kinematic
temperature, we concluded from the observed iron to nickel abundance
ratios that the abundances are non-solar.

\vspace{1cm}

We thank Craig Markwardt for careful reading of the manuscript.

\section*{Appendix A\@. Pile-up correction} 
The SIS/ASCA suffers pileup when observing a bright source
$\stackrel>\sim$ 100 counts s$^{-1}$.  If two or more photons fall on
a single pixel or neighboring pixels (Moore neighbor) on a CCD chip of
the SIS within the same readout period, they are recognized as a
single photon by the on-board event-detection algorithm, and their
energies are summed to make a single pulse height.  This phenomenon is
called ``pileup'' and causes a decrease in the photon counting rate
and hardening of the spectrum.  In observations of bright sources,
pileup is non-negligible and correction is necessary for precise
spectral analysis.  We have developed a new code\footnote{Included in
FTOOLS v5.0 available at
http://heasarc.gsfc.nasa.gov/docs/software/ftools} for this purpose.
This code also can cope with attitude fluctuations of the satellite,
which were in fact significant during the GRS 1915+105 observation in
1994.

Our code simulates the pileup using the observed events as the seeds.
First, the source position is determined on the CCD image without
correction, and the image core where the counting rate is higher than
a certain threshold is discarded from further analysis.  The threshold
was empirically determined to be 0.023 counts pixel$^{-1}$
readout$^{-1}$, above which we found the pileup effect was so
significant that uncorrectable.  The pixels on the CCD chip are
divided into several annuli centered at the source.  The annuli widths are
determined so that the counting rate per pixel within each annulus is
almost uniform.  The number of detected events in the $i$-th annulus and
the $j$-th readout is defined as $a_{ij}$.  By dividing the chips into
annuli, both the count rate and the image are eventually recovered.

Next, the temporal order of all the $N_i = \Sigma_j a_{ij}$ events in the
$i$-th annulus is randomized, while each event retains its position,
pulse-height, and charge-distribution pattern, i.e., grade
information.  The event series before and after randomization are
denoted as
\begin{equation}
\gamma_{i1},\; \gamma_{i2},\; \ldots \; ,\gamma_{i\,N_i}
\end{equation}
and
\begin{equation}
\gamma_{i\sigma(1)},\; \gamma_{i\sigma(2)},\; \ldots \; ,\gamma_{i\sigma(N_i)},
\end{equation}
where $\sigma(k)$ is a randomizing function which returns an integer
between 1 and $N_i$, and satisfies $\sigma(k) \neq \sigma(l)$ if $k
\neq l$.  For each read-out, the randomized events  are injected on a
virtual CCD chip, and the
same event detection algorithm as that on-board is applied (Gendreau 1995).
 For example, to simulate the $j$-th readout of the $i$-th ring, a series of events
\begin{equation}
\gamma_{i\sigma(\Sigma_{k=1}^{j-1}a_{ik}+1)},\;
\gamma_{i\sigma(\Sigma_{k=1}^{j-1}a_{ik}+2)},\; \ldots \;
,\gamma_{i\sigma(\Sigma_{k=1}^{j}a_{ik})}
\end{equation}
is created (note that $\Sigma_{k=1}^ja_{ik} - \Sigma_{k=1}^{j-1}a_{ik}
= a_{ij}$), and charge-distribution patterns on the virtual CCD chip
are calculated from the position, pulse-height, and original
charge-distribution pattern or grade of each input event. If the SIS
observation mode is BRIGHT and the information on split charge has
been lost, the charge is assumed to be 40 ADU (the split-threshold).
The spectrum of the injected events is recorded as $f_2(E)$.  If there
is no pileup, all $a_{ij}$ events should be detected.  On the other
hand, the detected events may be less than $a_{ij}$ if pileup is
accounted for.  In that case additional events are randomly picked up
from the total $N_i$ events and injected onto the virtual chip (also
added to $f_2(E)$) until $a_{ij}$ events are detected.  When the
number of detected events has reached $a_{ij}$, the spectrum
determined by the event detection algorithm is recorded as $f_3(E)$,
which will be the simulated spectrum affected by the pileup.  This
procedure is repeated for all the annuli and readouts.  It should be
noted that all the original events are used in the whole process at
least once.  By using all the events, fluctuations caused by random
pick up will be reduced to minimum.  Finally, the pure pileup spectrum
$f_4(E)$ and resultant pileup-corrected spectrum $f_5(E)$ are
estimated as
\begin{eqnarray}
f_4 & = & f_3 - f_2\\
f_5 & = & f_1 - f_4 =  f_1  + f_2 - f_3,
\end{eqnarray}
where $f_1(E)$ is the observed spectrum before the correction.  In
Fig.~\ref{kotani:fig:pileupspec}, an example of the pileup
correction is exhibited.  

\placefigure{kotani:fig:pileupspec}

\clearpage
\begin{table}
\caption{Observation log\label{kotani:tbl:obs}}
\begin{tabular}{ccccc}
\tableline
\tableline
No. &Start &End &  Exposure Time\tablenotemark{a} &SIS Mode\\ 
& (UT)& (UT)& (ks) &\\
\tableline
1 &1994/09/27 00:40 &1994/09/27  13:00 & 10 &BRIGHT\\
2 &1995/04/20 09:50 &1995/04/20  22:40 & 16 &FAINT\\
3 &1996/10/23 02:40 &1996/10/23  18:20 & 11 &BRIGHT\\
4 &1997/04/25 08:10 &1997/04/25  21:00 & 20 &BRIGHT\\
5 &1998/04/04 16:20 &1998/04/05  03:30 & 20 &BRIGHT\\
6 &1999/04/15 20:10 &1999/04/16  15:00 & 20 &BRIGHT\\
\tableline
\end{tabular}
\tablenotetext{a}{Bit HIGH only.}
\end{table}

\begin{table}
\caption{Best-fit parameters of the GIS spectral continuum\label{kotani:tbl:bestfitgis}}
\begin{tabular}{cccccccc}
\tableline
\tableline
No.     &$N_{\rm H1}$   &$f$    &$N_{\rm H2}$   &$A$\tablenotemark{a}       &$E_{\rm fold}$ &$\Gamma$
&$\chi^2_\nu$ ($\nu$)\\
        &($10^{22}$ cm$^{-2}$) & &($10^{22}$ cm$^{-2}$) & &(keV)\\
\tableline
1\tablenotemark{b} &$2.97^{+0.10}_{-0.10}$ &$0.551^{+0.040}_{-0.040}$ &$55^{+6}_{-5}$
&$4.67^{+0.44}_{-0.41}$ &$1.28^{+0.04}_{-0.04}$ &$-0.91^{+0.13}_{-0.13}$ &1.13 (131)\\
2\tablenotemark{b} &$3.46^{+0.10}_{-0.10}$ &$0.563^{+0.065}_{-0.070}$ &$81^{+10}_{-10}$
&$5.84^{+0.69}_{-0.60}$ &$1.35^{+0.05}_{-0.05}$ &$-1.18^{+0.13}_{-0.13}$ &1.13 (131)\\
3 &$4.75^{+0.07}_{-0.04}$ &$0.219^{+0.066}_{-0.045}$ &$71^{+22}_{-14}$
&$17.6^{+1.4}_{-1.2}$ &$4.61^{+0.28}_{-0.27}$ &$1.50^{+0.07}_{-0.08}$ &1.12 (174)\\
\tableline
\end{tabular}
\tablenotetext{a}{In unit of photons s$^{-1}$ cm$^{-2}$ keV$^{-1}$ at 1 keV.}
\tablenotetext{b}{The Energy range 6--8 keV was excluded.}
\end{table}

\begin{table}
\caption{Best-fit parameters of the feature around iron-K absorption
\label{kotani:tbl:bestfitsis}}
\begin{tabular}{crcc}
\tableline\tableline
Model Component &Parameter              &1994   &1995\\
\tableline
Emission Line   &Energy (keV)   &$6.484^{+0.044}_{-0.045}$      &$\cdots$\\
                &Width  (eV)    & 0 (fixed)     &$\cdots$\\
                &EW (eV)        &$15.8^{+6.3}_{-6.9}$   &$\cdots$\\
Negative Gaussian A     &Energy (keV)   &$6.720^{+0.015}_{-0.017}$
&$6.690^{+0.028}_{-0.034}$\\
                &Width\tablenotemark{a} (eV) &$<39$  &$<89$\\
                &EW (eV)        &$35.8^{+3.7}_{-7.8}$   &$27.5^{+4.7}_{-6.3}$\\
Negative Gaussian B     &Energy (keV)   &$7.000^{+0.020}_{-0.014}$
&$6.964^{+0.012}_{-0.024}$\\
                &EW (eV)        & $43.2^{+6.3}_{-5.3}$  &$48.3^{+5.4}_{-5.7}$\\
Negative Gaussian C     &Energy (keV)   &$7.843^{+0.037}_{-0.035}$
&$7.839^{+0.043}_{-0.037}$\\
                &EW (eV)        &$30.8^{+11.2}_{-5.8}$  &$29.9^{+8.4}_{-8.5}$\\
Negative Gaussian D     &Energy (keV)   &$8.199^{+0.065}_{-0.064}$    &$8.181^{+0.060}_{-0.057}$\\
                &EW (eV)        &$22^{+11}_{-12}$    &$19^{+9}_{-10}$\\
Absorption Edge &Energy (keV)   &$\cdots$    & 7.117 (fixed)\\
                &Max.\ Opt.\ Depth      &$\cdots$    &$0.122^{+0.027}_{-0.027}$\\
\tableline
$\chi^2_\nu$ ($\nu$)    &       &1.03 (56)      &1.05 (57)\\
\tableline
\end{tabular}
\tablenotetext{a}{Upper limit at 1 $\sigma$ confidence level.  Set free but same for all negative Gaussians.}
\end{table}

\begin{table}
\caption{Best-fit parameters of the absorption features at 4 keV\label{kotani:tbl:bestfitsisca}}
\begin{tabular}{crcc}
\tableline\tableline
Model Component &Parameter              &1994   &1995\\
\tableline
Negative Gaussian X     &Energy (keV)   &$4.071^{+0.068}_{-0.097}$
&$4.05^{+0.11}_{-0.11}$\\
                &Width (eV)     &$<210$ &10 (fixed)\\
                &EW (eV)        &$10.2^{+7.5}_{-5.4}$   &$<7.9$\\
\tableline
$\chi^2_\nu$ ($\nu$)    &       &0.80 (56)      &1.24 (57)\\
\tableline
\end{tabular}
\end{table}

\begin{table}
\caption{Absorption-line candidates\label{kotani:tbl:candidates}}
\begin{tabular}{ccccc}
\tableline\tableline
Ion     &Upper Level    &Energy         &Oscillator Strength &Einstein Coefficient\\
        &       &(keV)  &       &(s$^{-1}$)\\
\tableline
Fe {\sc xxv}    &$1s^12p^1$ $\rm^3P_1$      &6.6684925 (1)       &0.0687 (1)
&$4.42\times10^{13}$\\  
Fe {\sc xxv}    &$1s^12p^1$ $\rm^1P_1$  &6.7011266 (1)  &0.703 (1)
&$4.57\times10^{14}$\\  
Fe {\sc xxvi}   &$2p^1$ $\rm^2P_{1/2,3/2}$      &6.9519639, 6.9731781 (3) &0.44 (4)
&$1.5\times10^{14}$\\   
Ni {\sc xxvii}  &$1s^12p^1$ $\rm^3P_1$      &7.7668938 (1)      &0.0883 (1)
&$7.70\times10^{13}$\\  
Ni {\sc xxvii}  &$1s^12p^1$ $\rm^1P_1$  &7.8062340 (1)  &0.683 (1)
&$6.02\times10^{14}$\\  
Fe {\sc xxv}    &$1s^13p^1$ $\rm^1P_1$  &7.8820244 (2)\        &0.138 (2)
&$1.24\times10^{14}$\\  
Ni {\sc xxviii} &$2p^1$ $\rm^2P_{1/2,3/2}$      &8.0731039, 8.1017429 (3) &0.44 (4)
&$2.07\times10^{14}$\\  
Fe {\sc xxvi}   &$3p^1$ $\rm^2P_{1/2,3/2}$      &8.2636944, 8.2526875 (3) &0.046 (4)
&$2.3\times10^{13}$\\   
\tableline
\end{tabular}
\tablerefs{(1) Drake  1979;
 (2) Lin, Johnson, \& Dalgarno 1977;
 (3) Sugar \& Corliss 1985;
 (4) Mewe, Gronenschild, \& van den Oord 1985, and references therein.}
\end{table}

\begin{table}
\caption{Column density of absorbing ions\label{kotani:tbl:column}}

\begin{tabular}{ccccc}
\tableline\tableline
        &\multicolumn{4}{c}{High-Temperature Limit ($kT = 1000$ keV)}\\
\tableline
        &Fe {\sc xxv}    &Fe  {\sc xxvi}  &Ni  {\sc xxvii} &Ni {\sc xxviii}\\
1994    &$17.7^{+0.1}_{-0.1}$    &$18.0^{+0.1}_{-0.1}$    &$17.5^{+0.2}_{-0.2}$    &$17.6^{+0.3}_{-0.7}$\\
1995   &$17.6^{+0.1}_{-0.1}$    &$18.1^{+0.1}_{-0.1}$   &$17.5^{+0.2}_{-0.2}$
&$17.4^{+0.5}_{-0.9}$\\
\tableline\tableline
        &\multicolumn{4}{c}{Low-Temperature Limit ($kT = 0.1$ keV)}\\
\tableline
        &Fe {\sc xxv}    &Fe  {\sc xxvi}  &Ni  {\sc xxvii} &Ni {\sc xxviii}\\
1994    &$19.3^{+0.1}_{-0.2}$    &$20.3^{+0.1}_{-0.1}$    &$18.8^{+0.4}_{-0.3}$    &$19.4^{+0.5}_{-1.2}$\\
1995    &$19.0^{+0.1}_{-0.2}$   &$20.4^{+0.1}_{-0.1}$   &$18.8^{+0.3}_{-0.4}$
&$19.1^{+0.8}_{-1.5}$\\
\tableline
\end{tabular}
\tablecomments{In units of 
log$_{10}$[cm$^{-2}$].}
\end{table}

\begin{table}
\caption{Parameters of the photo-ionized plasma
\label{kotani:tbl:photo}}
\begin{tabular}{ccccccc}
\tableline\tableline 
&$L$    &$\xi_0$ &$r_0$\tablenotemark{a} &$\xi_1$ &$r_1$\tablenotemark{a} &$n\rm_p$\\
&(erg s$^{-1}$)&(erg s$^{-1}$ cm)&(cm) &(erg s$^{-1}$ cm)&(cm) &(cm$^{-3}$) \\
\tableline
1994 &$4.0\times10^{38}$ (fixed) &$10^{5.5}$ (fixed) &$9\times10^{10}$ &$10^{3.8}$
&$6\times10^{11}$ &$1.5\times10^{11}$\\
1995 &$8.8\times10^{38}$ (fixed) &$10^{5.5}$ (fixed) &$14\times10^{10}$ &$10^{3.8}$
&$9\times10^{11}$ &$1.5\times10^{11}$\\
\tableline
\end{tabular}
\tablecomments{Subscript``0'' and ``1'' denote values at the
innermost and outermost boundary,
respectively.}
\tablenotemark{a}{Calculated from ionization parameter  $\xi$, and not
a free parameter.}
\end{table}

\begin{table}
\caption{Model prediction of ion column density
\label{kotani:tbl:calccolumn}}

\begin{tabular}{ccccccccccc}
\tableline\tableline
        &H {\sc ii}     &O {\sc viii} &Si {\sc xiv} &S {\sc xvi} &Ca {\sc xx} &Fe {\sc xxv} &Fe {\sc xxvi}
&Ni  {\sc xxvii} &Ni {\sc xxviii}\\
\tableline
1994 &22.9 &16.1 &15.9 &16.0 &15.9 &17.7 &18.0 &15.1 &16.7\\
1995 &23.1 &16.4 &16.0 &16.1 &15.9 &17.6 &18.1 &16.7 &16.9\\
\tableline
\end{tabular}
\tablecomments{In unit of log$_{10}$[cm$^{-2}$].  Ion species less
than $10^{16}$ cm$^{-2}$ were omitted, as were naked ions.
The abundance was assumed to be solar.  Input parameters were
determined so that Fe {\sc xxv} and Fe {\sc xxvi} were consistent with
the observations.}
\end{table}

\clearpage

\begin{figure}
\begin{minipage}{0.45\textwidth}
\resizebox{7cm}{!}{\rotatebox{270}{\includegraphics{ew.fe.25.a.ps}}}
(a) Fe {\sc xxv} K$\alpha$
\end{minipage}
\begin{minipage}{0.45\textwidth}
\resizebox{7cm}{!}{\rotatebox{270}{\includegraphics{ew.fe.26.a.ps}}}
(b) Fe {\sc xxvi} K$\alpha$
\end{minipage}
\begin{minipage}{0.45\textwidth}
\resizebox{7cm}{!}{\rotatebox{270}{\includegraphics{ew.fe.25.b.ps}}}
(c) Fe {\sc xxv} K$\beta$
\end{minipage}
\begin{minipage}{0.45\textwidth}
\resizebox{7cm}{!}{\rotatebox{270}{\includegraphics{ew.fe.26.b.ps}}}
(d) Fe {\sc xxvi} K$\beta$
\end{minipage}
\begin{minipage}{0.45\textwidth}
\resizebox{7cm}{!}{\rotatebox{270}{\includegraphics{ew.ni.27.a.ps}}}
(e) Ni {\sc xxvii} K$\alpha$
\end{minipage}
\begin{minipage}{0.45\textwidth}
\resizebox{7cm}{!}{\rotatebox{270}{\includegraphics{ew.ni.28.a.ps}}}
(f) Ni {\sc xxviii} K$\alpha$
\end{minipage}

\caption{Curves of growth.  (a) Fe {\sc xxv} K$\alpha$, (b) Fe {\sc
xxvi} K$\alpha$, (c) Fe {\sc xxv} K$\beta$, (d) Fe {\sc xxvi}
K$\beta$, (e) Ni {\sc xxvii} K$\alpha$, and (f) Ni {\sc xxviii}
K$\alpha$.  The curves in each panel correspond to kinematic
temperature of 1000 keV, 100 keV, 10 keV, 1 keV, and 0.1 keV, from
high to low.
\label{kotani:fig:curve}}
\end{figure}

\begin{figure}
\includegraphics{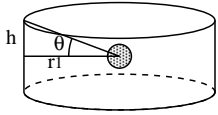}

\caption{Assumed geometry.  An X-ray source (gray sphere) is embodied
in a plasma disk expressed as a frame.
\label{kotani:fig:geometry}}
\end{figure}

\begin{figure}
\includegraphics{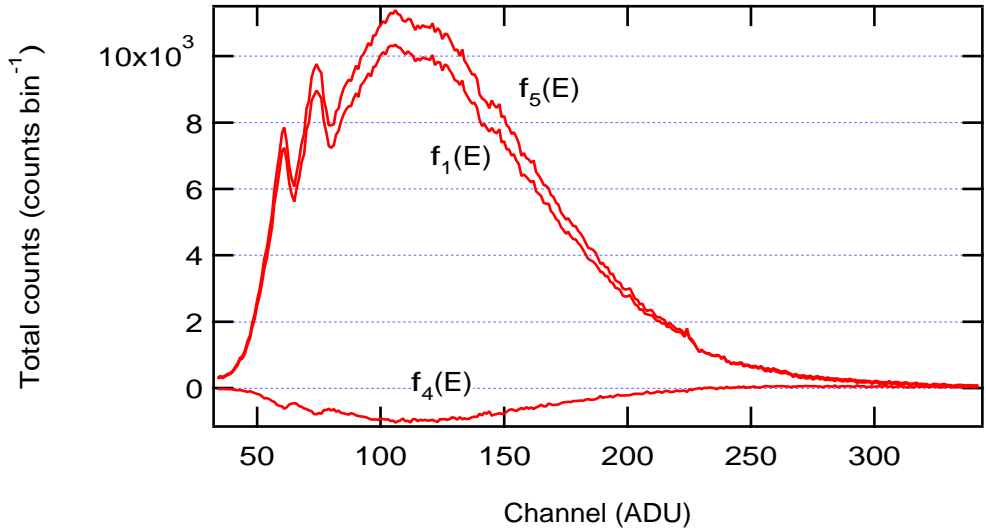}

\caption{Spectrum before and after pile-up correction.  
The Sensor-0 Chip-1 data of 1994 are presented as an
example.  The observed spectrum before correction $f_1(E)$, the
calculated pure pileup component $f_4(E)$, and the resultant
pileup-corrected spectrum $f_5(E)$ are shown.  Energy is expressed in
analog-to-digital-converter units (ADU), and the range is between
0.5--10 keV\@.  In the lower energy band (50--200 ADU), photons are lost by pileup
and thus $f_4(E)$ is negative by definition.  In the higher energy
band ($>$ 250 ADU), pseudo photons are made by pileup.  Although it is difficult to
discern by eye in a linear plot, the effect in the higher band is not
negligible.
\label{kotani:fig:pileupspec}}
\end{figure}

\clearpage

\begin{center}
{\large \bf Erratum:  ``ASCA Observations of the Absorption-Line Features from the
Super-Luminal Jet Source GRS 1915+105'' (ApJ, 539, 413 [2000])}
\end{center}

We have discovered errors in the calculation of some of the Einstein
coefficients in Table~5 and some plots in Figure~3.  Due to the errors,
the square-root section of the curves of growth of Fe {\sc xxvi}
K$\alpha$ and K$\beta$, and Ni {\sc xxviii} K$\alpha$ in Figure~3 were a
few times underestimated.  We correct Table~5 and re-plot Figure~3
together with the unaffected curves\footnote{The program to calculate
the corrected curves of growth is available at {\tt
http://www.hp.phys.titech.ac.jp/kotani/cog/index.html}}.  We thank Aya Kubota
and Chirs Done for pointing out this error.

\clearpage

\begin{table}
\begin{center}
TABLE~5\\
Absorption-line candidates
\begin{tabular}{ccccc}
\tableline\tableline
Ion     &Upper Level    &Energy         &Oscillator Strength &Einstein Coefficient\\
        &       &(keV)  &       &(s$^{-1}$)\\
\tableline
Fe {\sc xxv}    &$1s^12p^1$ $\rm^3P_1$      &6.6684925 (1)       &0.0687 (1)
&$4.42\times10^{13}$\\  
Fe {\sc xxv}    &$1s^12p^1$ $\rm^1P_1$  &6.7011266 (1)  &0.703 (1)
&$4.57\times10^{14}$\\  
Fe {\sc xxvi}   &$2p^1$ $\rm^2P_{1/2,3/2}$      &6.9519639, 6.9731781 (3) &0.44 (4)
&$3.1\times10^{14}$\tablenotemark{a}\\   
Ni {\sc xxvii}  &$1s^12p^1$ $\rm^3P_1$      &7.7668938 (1)      &0.0883 (1)
&$7.70\times10^{13}$\\  
Ni {\sc xxvii}  &$1s^12p^1$ $\rm^1P_1$  &7.8062340 (1)  &0.683 (1)
&$6.02\times10^{14}$\\  
Fe {\sc xxv}    &$1s^13p^1$ $\rm^1P_1$  &7.8820244 (2)\        &0.138 (2)
&$1.24\times10^{14}$\\  
Ni {\sc xxviii} &$2p^1$ $\rm^2P_{1/2,3/2}$      &8.0731039, 8.1017429 (3) &0.44 (4)
&$4.1\times10^{14}$\tablenotemark{a}\\  
Fe {\sc xxvi}   &$3p^1$ $\rm^2P_{1/2,3/2}$      &8.24636944\tablenotemark{b}, 8.2526875 (3) &0.046 (4)
&$4.5\times10^{13}$\tablenotemark{a}\\   
\tableline
\end{tabular}
\end{center}
\tablenotetext{a}{Corrected.}
\tablenotetext{b}{Typo corrected.}
\tablerefs{(1) Drake, G. W. F. \pra, 19, 1387 [1979];
 (2) Lin, C. D., Johnson, W. R., \& Dalgarno, A. \apj, 217, 1011 [1977];
 (3) Sugar, J.,  \& Corliss,  C. J. Phys.\ Chem.\ Ref.\ Data, 14,
  Supplement No. 2 [1985];
 (4) Mewe, R., Gronenschild, E. H. B. M., \& van den Oord, G. H. J. 
A\&AS, 62, 197 [1985], and references
  therein}
\end{table}

\clearpage

\begin{figure}
\begin{centering}
\includegraphics[angle=270,scale=.33]{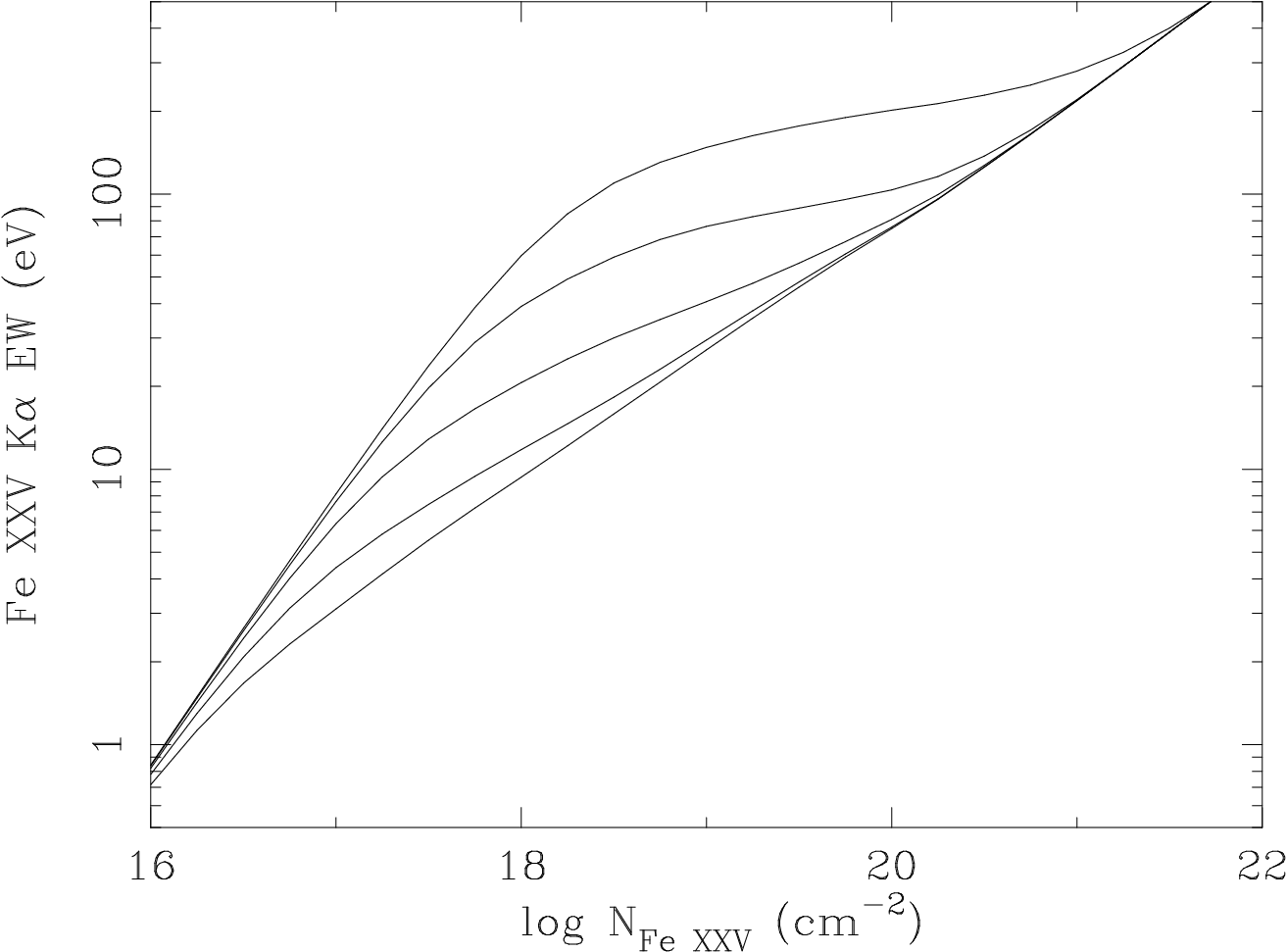}
\includegraphics[angle=270,scale=.33]{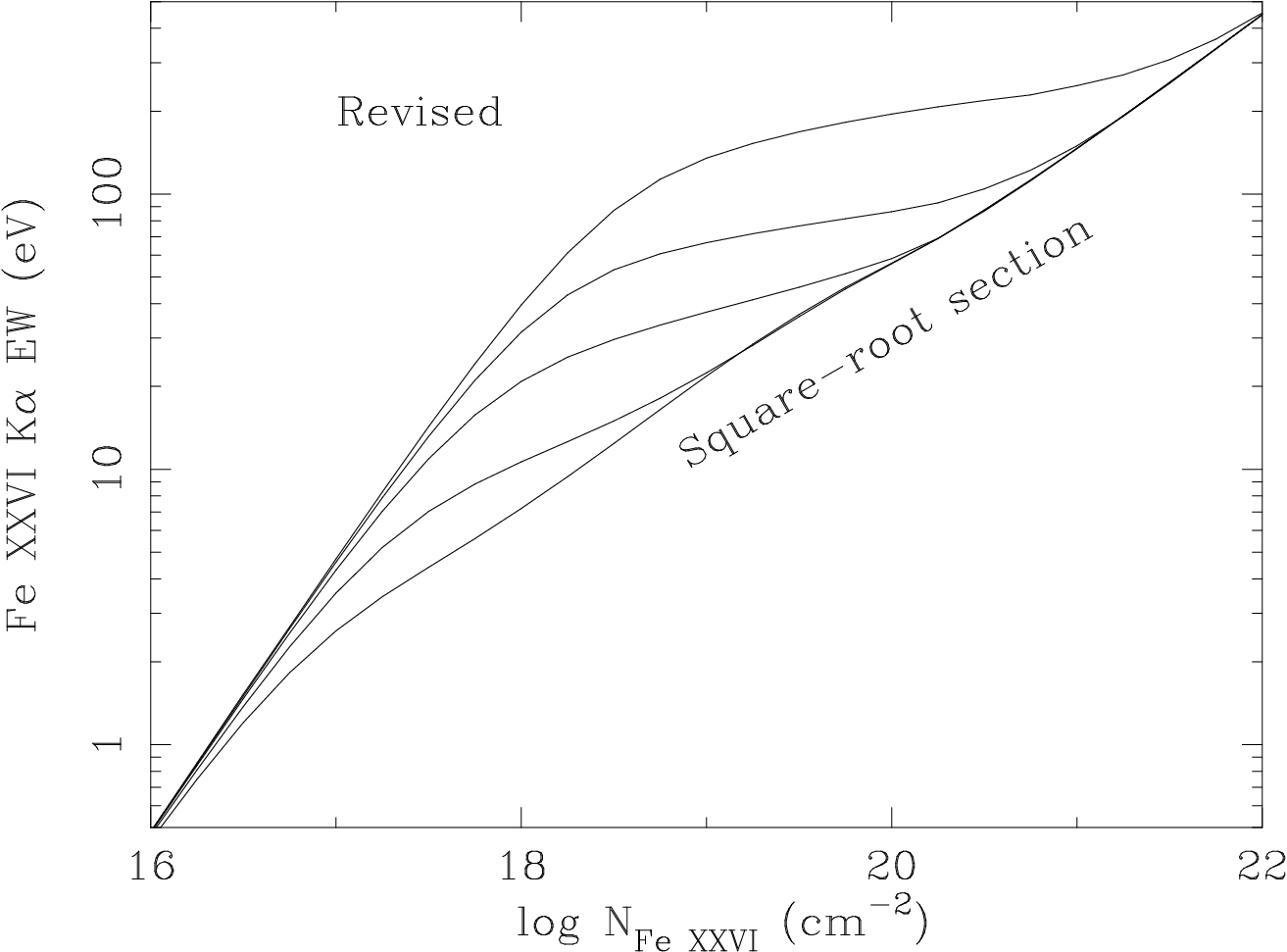}
Fig.~3a~~~~~~~~~~~~~~~~~~~~~~~~~~~~~~~~~~~~~~Fig.~3b

\includegraphics[angle=270,scale=.33]{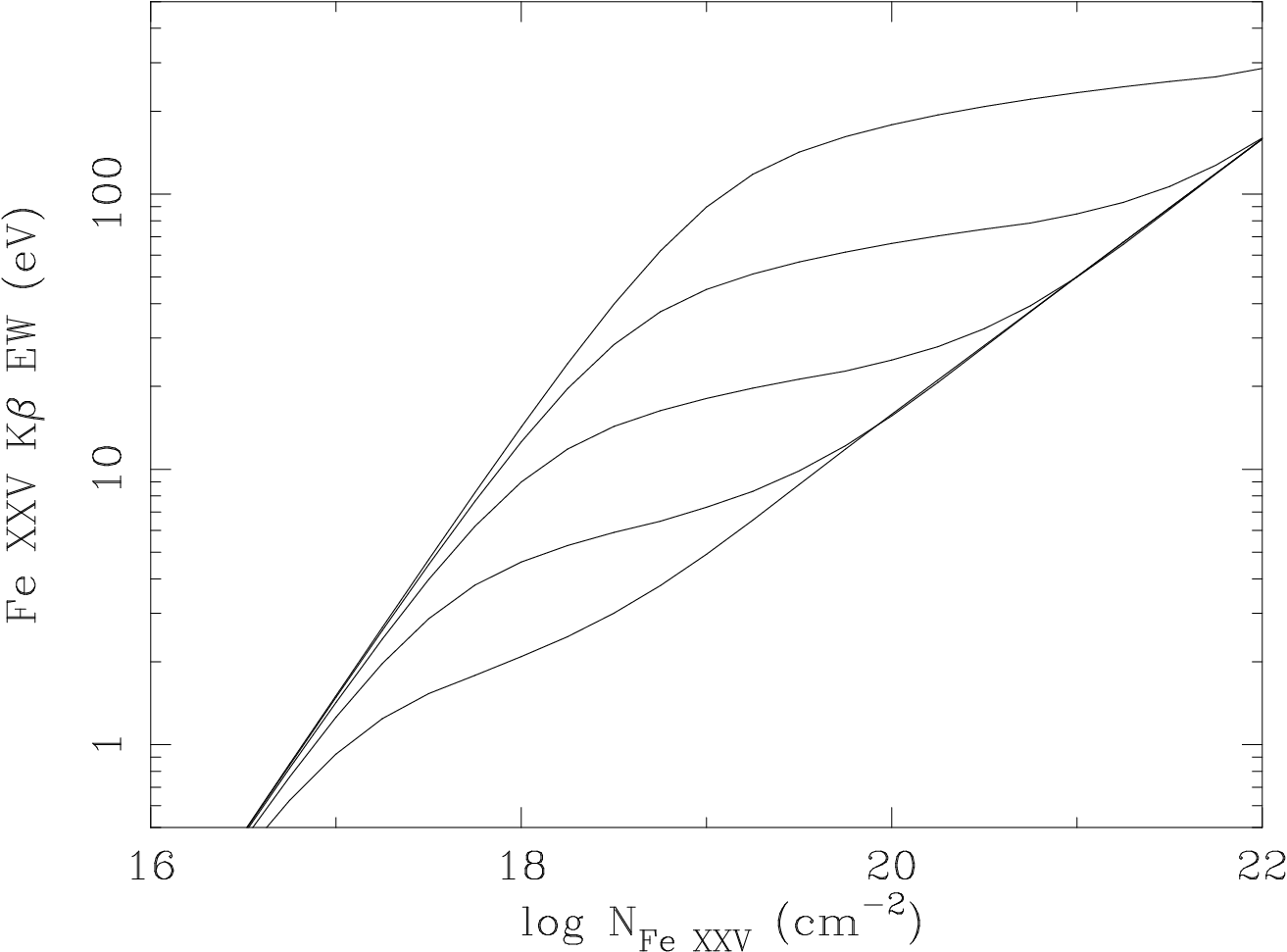}
\includegraphics[angle=270,scale=.33]{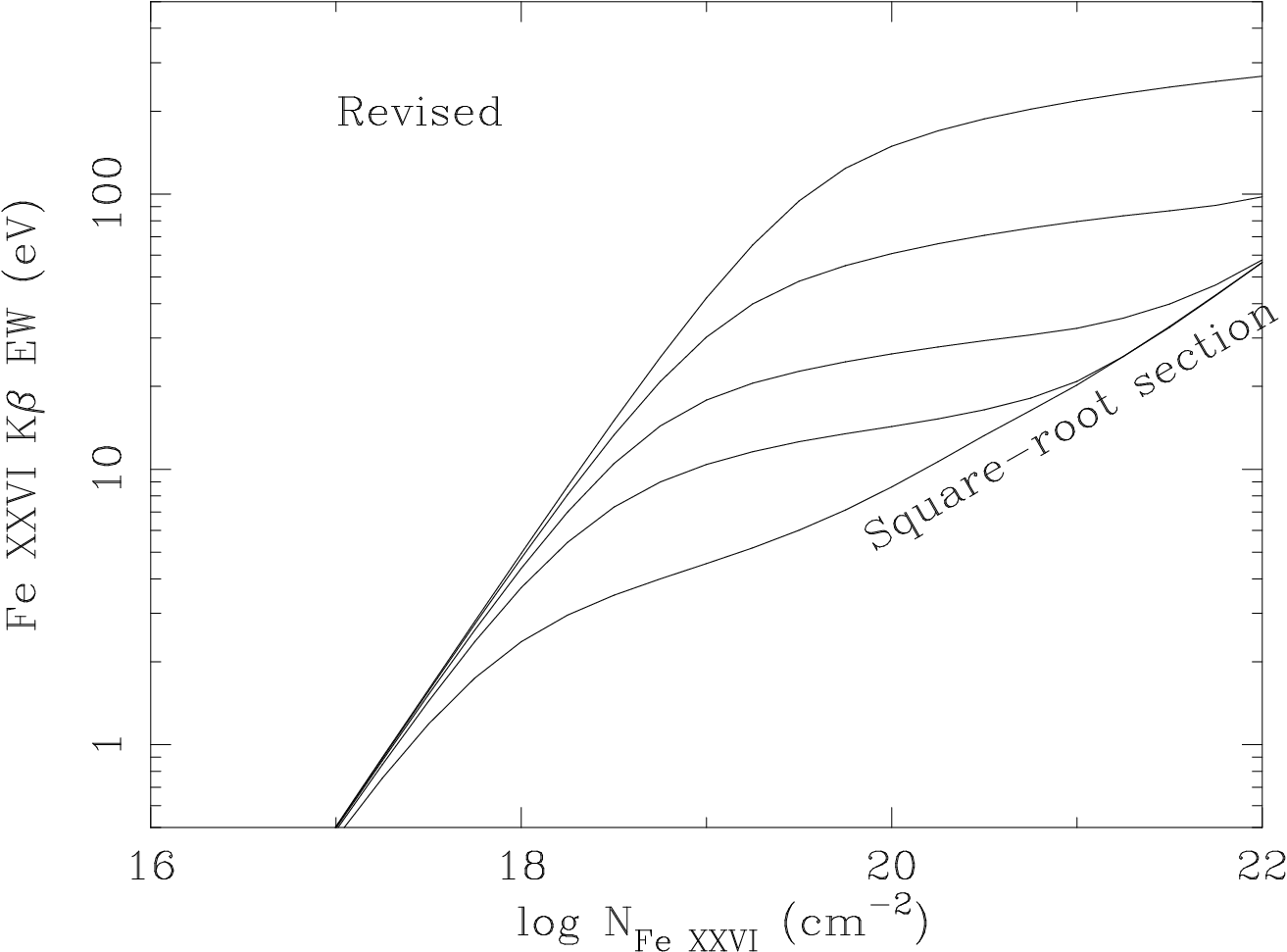}
Fig.~3c~~~~~~~~~~~~~~~~~~~~~~~~~~~~~~~~~~~~~~Fig.~3d

\includegraphics[angle=270,scale=.33]{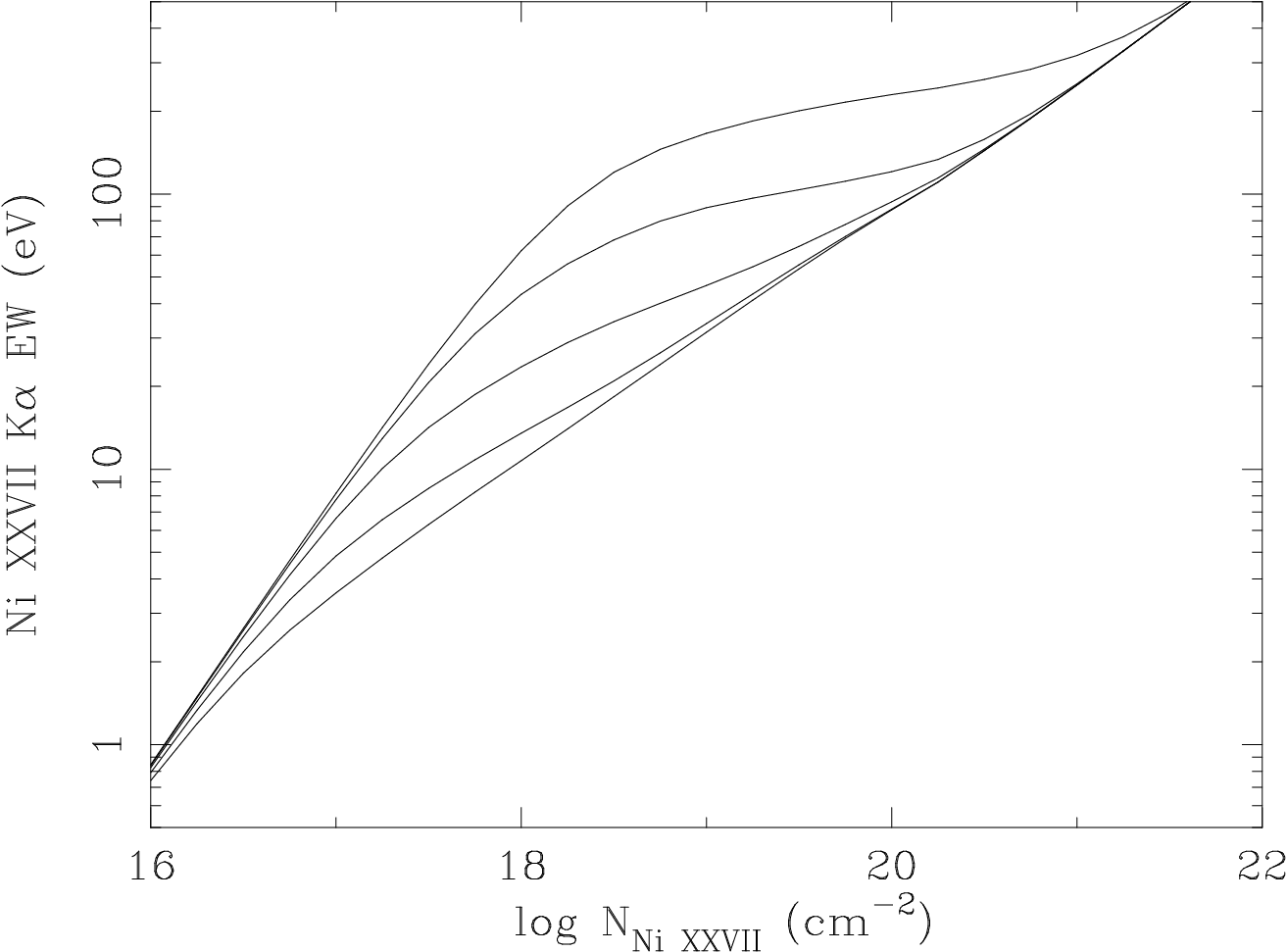}
\includegraphics[angle=270,scale=.33]{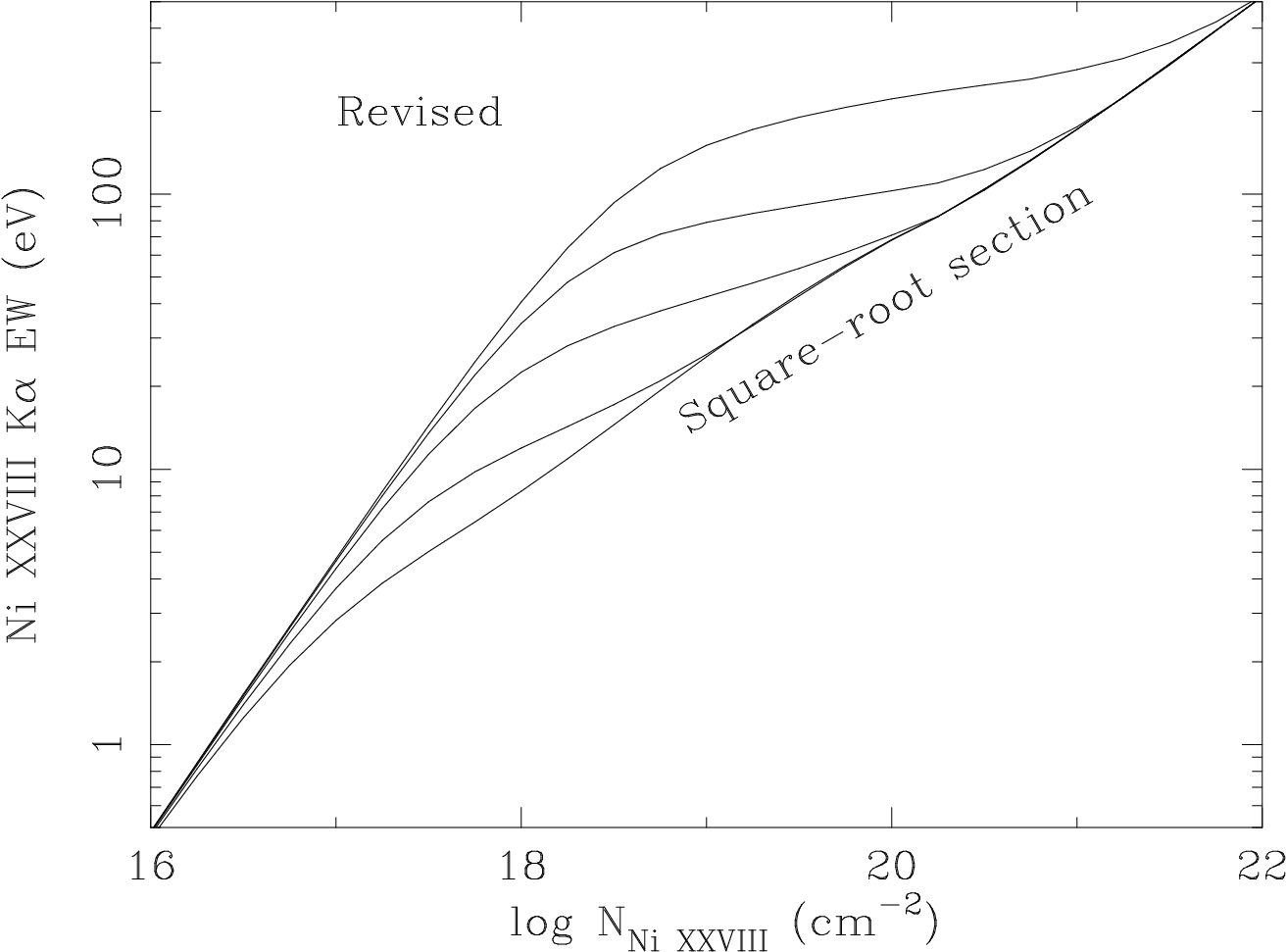}
Fig.~3e~~~~~~~~~~~~~~~~~~~~~~~~~~~~~~~~~~~~~~Fig.~3f

\end{centering}

Fig.~3---Curves of growth.  (a) Fe {\sc xxv} K$\alpha$, (b) Fe {\sc
xxvi} K$\alpha$, (c) Fe {\sc xxv} K$\beta$, (d) Fe {\sc xxvi}
K$\beta$, (e) Ni {\sc xxvii} K$\alpha$, and (f) Ni {\sc xxviii}
K$\alpha$.  The curves in each panel correspond to kinematic
temperature of 1000 keV, 100 keV, 10 keV, 1 keV, and 0.1 keV, from
high to low.

This figure replaces Fig.~3 of the original manuscript.  The square-root
section of the curves in Fig.~3b, 3d, and 3f are a few times increased
by this correction.  The unaffected figures, 3a, 3c, and 3e are also shown.
\end{figure}

\end{document}